# New Aspects of Photocurrent Generation at Graphene pn Junctions Revealed by Ultrafast Optical Measurements

Dong Sun[1], Grant Aivazian[1], Aaron M. Jones[1], Wang Yao[2], David Cobden[1], Xiaodong Xu[1,3*]

1. Department of Physics, University of Washington, Seattle, Washington 98195, USA
2. Department of Physics and Center of Theoretical and Computational Physics, The University of Hong Kong, Hong Kong, China
3. Department of Material Science and Engineering, University of Washington, Seattle, Washington 98195, USA

Email: xuxd@uw.edu

**Introduction**: The unusual electrical[1,2,3,4,5] and optical properties[6] of graphene make it a promising candidate for optoelectronic applications[7,8,9,10,11]. An important, but as yet unexplored aspect is the role of photo-excited hot carriers in charge and energy transport at graphene interfaces[12,13,14]. Here, we perform time-resolved (~250 fs) scanning photocurrent microscopy on a tunable graphene pn junction. The ultrafast pump-probe measurements yield a photocurrent response time of ~1.5 ps at room temperature increasing to ~4 ps at 20 K. Combined with the negligible dependence of photocurrent amplitude on environmental temperature this implies that hot carriers rather than phonons dominate energy transport at high frequencies. Gate-dependent pump-probe measurements demonstrate that both thermoelectric[15] and built-in electric field[16] effects contribute to the photocurrent excited by laser pulses. The relative weight of each contribution depends on the junction configuration. A single laser beam excitation also displays multiple polarity-reversals as a function of carrier density, a signature of impact ionization[13]. Our results enhance the understanding of non-equilibrium electron dynamics, electron-electron interactions, and electron-phonon interactions in graphene. They also determine fundamental limits on ultrafast device operation speeds (~500 GHz) for potential graphene-based photon detection, sensing, and communication.[7,8,10,11]

**Text**:
Graphene's combination of high electron mobility[1,2,3,4,5] and broadband absorption[6] put it at center stage for new optoelectronic and photonic applications[10], including transparent electrodes[17], mode-locked lasers[18], and high-speed optical modulators[11]. A first step of particular interest is to develop tunable, high speed, and broadband graphene photodetectors[7,8,12,13,14,15,16,19,20,21]. Previous work has demonstrated a detection bandwidth of ~40 GHz for a graphene/metal contact (GM) device[7]. Simple analysis suggests an operation bandwidth of ~500 GHz[7]. On the other hand, despite the successful demonstration of a prototype GM photodetector, basic questions related to the photocurrent generation mechanism remain, such as whether the built-in field or the photo-thermoelectric effect dominates the photocurrent generation[7,12,13,14,15,16,21,22], and the role of hot-carrier transport in ultrafast devices and possible carrier multiplication processes[13]. This is mainly due to the speed limitations of electrical measurements and to the use of continuous wave (CW) excitation[7,9,11], in which the electron temperature ($T_e$) is close to the lattice temperature ($T_L$). For picosecond time scale operation, nonequilibrium effects come into play since the hot carriers do not have time to reach thermal equilibrium with the lattice. Moreover, even under CW excitation, due to the relatively slow (~10 ns) acoustic phonon scattering rate, recent theory indicates that hot-carrier transport and carrier multiplication[13] could dominate photocurrent generation. Therefore, it is important to study the hot-carrier transport with sub-ps temporal resolution experimentally.

To do this, we apply spatially (~1 μm) and temporally (~ 250 fs) resolved scanning photocurrent microscopy to investigate tunable graphene pn junctions. We utilize a dual-gate structure which allows independent control of the charge density in adjacent regions[12]. We have studied both single- and few-layer exfoliated graphene as well as single-layer graphene grown by chemical vapor deposition. Measurements on the latter show similar behavior and are presented in the Supplementary Materials. The data presented here are all taken from a selected exfoliated trilayer device. Figure 1a is a schematic diagram of the measurement setup, and an optical microscope image of the device is shown on the left in Fig. 1b. Application of voltages to the top and bottom gate creates a graphene homojunction on either side of the top gate which can be tuned between pp, pn, np and nn. Here the first and second letters represent the doping outside and underneath the top gate, respectively. Since the fabrication process results in a positive doping which shifts the neutrality point to a back gate voltage of ~70 V we define the effective back gate voltage $V_{bg}$ to be measured relative to this value. Figure 1c shows resistance as a function of $V_{bg}$ and top gate voltage $V_{tg}$. The tilted dashed line indicates where the graphene under the top gate is at the neutrality point.

First, we perform standard photocurrent microscopy by scanning the focused pulsed laser over the device. The results of such a measurement at a temperature of 20 K and laser power of ~70 μW, or ~4×10$^{14}$ photons/cm$^2$ per pulse are shown in Fig. 1b. Photocurrent is observed at the graphene-metal contact interfaces and at the pn junctions. We focus on the latter. We first consider the results of excitation by a single laser beam. Figure 1d shows the photocurrent as a function of gate voltages with the laser located at the pn junction indicated by the circle in Fig. 1b. Fig. 1e is a single line trace at $V_{bg}$=30 V, which corresponds to global n-doping. We observe that the photocurrent crosses zero twice as $V_{tg}$ varies. This is a clear indication of photocurrent generation from the thermoelectric effect. The built-in field effect only changes sign once as a function of $V_{tg}$, and only the addition of the thermoelectric effect can account for multiple zero crossings[12].

Next, we apply ultrafast pump-probe techniques to investigate the hot-carrier dynamics. Two independently controlled pulsed beams are focused at the same spot. The pump and probe pulses are cross-polarized to minimize interference near zero delay. The probe beam is chopped and a lock-in amplifier is used to detect the resulting modulation of the photocurrent. Hence, the measured signal can be thought of as the probe-induced photocurrent, which we refer to as the PC, a quantity whose sensitivity to the presence of the pump pulse can be investigated as a function of delay and pulse amplitude (see Supplementary Materials for details).

Figure 2a shows typical pump-probe measurements at a temperature of 20 K. The three datasets correspond to different pump powers. Data have been normalized to the PC generated in the absence of the pump, and all pump-probe data are for a probe power of 70 μW unless otherwise specified. The presence of the pump only affects the PC near zero delay, producing a sharp dip which has the appearance of a saturation effect. The response time τ, defined as the half-width half-maximum of the dip, is ~ 4 ps, with no clear dependence on pump power (red points in Fig. 2b). The PC at zero delay is plotted as a function of pump power in Fig. 2b (black squares). The decrease of the PC as pump power increases is also consistent with a saturation effect, as is the saturation of the PC as a function of probe power with no pump, as shown in Fig. 2c. The latter has approximate dependence $I_{pc} \sim P^{0.7}$ on probe power $P$.

The dynamics change significantly with environmental temperature, as can be seen in Fig. 3a which shows measurements at 250 K and 20 K. We find that $\tau$ increases from ~1.5 ps to ~4 ps on cooling from 295 K to 20 K (Fig. 3b). Interestingly, over the same temperature range the PC generated by CW excitation increases by a factor of ~13 while the PC generated by pulsed excitation at the same power is almost unchanged (Fig. 3c).

We now discuss the implications of these findings. It is known that the pulse-excited electrons equilibrate amongst themselves on a timescale of tens of femtoseconds by electron-electron interactions, after which the electronic system can be characterized by a local electron temperature.[23,24,25,26] This temperature, $T_e$, determines the PC generated by both the built-in field and photo-thermoelectric effects. Understanding the details of the saturation effect requires a knowledge of the specific heats of electrons and optical phonons and of the thermoelectric coefficient at elevated $T_e$, which is not presently available. However, roughly speaking, saturation implies that the PC increases more slowly with $T_e$ than $T_e$ increases with laser power, ie, if $I_{pc} \sim T_e^\alpha$ and $T_e \sim P^{1/\beta}$ so that $I_{pc} \sim P^{\alpha/\beta}$ then $\alpha < \beta$ (see Supplementary Materials.).

Pulsed excitation is significantly different from CW excitation. CW excitation results in a steady, low $T_e$[15,27]. Under pulsed excitation $T_e$ spikes up following each pulse, corresponding to the generation of hot carriers, which then rapidly cool before the next pulse arrives. This cooling involves two mechanisms. The first is optical phonon-mediated cooling, which reduces $T_e$ to ~400 K on a timescale of several picoseconds[25,26,28,29] after which further cooling occurs through the slow coupling to acoustic phonons on time scales of ~100 ps to 10 ns (depending on carrier density)[24,28,29].

The second, which is not relevant to pump-probe measurements of isolated graphene sheets[25,26,28,29] is energy transport away from the junction by the carriers. First, the hot carriers drift out of the junction under the built-in electric field. This produces cooling on a time scale of ~100 ps in our device (Supplementary Materials). Second, the Peltier effect cools the junction because of the large change in Seebeck coefficient across the junction. Numerical simulations show that the PC decays on a time scale of ~ 5 ps through this mechanism (Fig. S8). The Peltier cooling rate is proportional to the photo-thermoelectric current which is inversely proportional to the RC time constant of the circuit. This suggests that $\tau$ should increase as the device resistance increase: indeed, the two appear to be correlated, as shown in Figure 3b. Hence the evidence points to Peltier cooling as playing an important role in hot-carrier relaxation at the pn junction.

Unlike coupling to optical phonons, which cools the whole laser excitation area, the Peltier mechanism only cools the junction region and hence the two mechanisms may be distinguished in future studies using high resolution techniques. We note that Peltier cooling has already been found to play a significant role in cooling of graphene-metal contacts using scanning thermal microscopy[30].

We can use the above transient carrier transport picture to understand the dramatically different temperature dependences of the PC amplitude under pulsed and CW excitations apparent in Fig. 3c. Such a temperature dependence under CW excitation has been reported previously and can be explained well in terms of acoustic phonon-mediated cooling[15]. In the pulsed case, due to the fast (~ps) cooling through optical phonons and the Peltier effect, acoustic phonons with their much longer scattering times are not involved and hence the lattice temperature does not affect the PC generation.

Finally, we turn to the measurements of the PC as a function of top and bottom gate voltages, for example those shown in Fig. 4a made at a series of $V_{tg}$ with $V_{bg}$ = -24.2 V. Under most conditions the PC shows saturation behavior, i.e., it is reduced in the presence of the pump. This is illustrated in Fig.

4b, where at $V_{bg}$ = -44.2 V the PC with the pump at zero delay (red trace) appears to be a suppressed version of that with no pump (black trace). Remarkably however, with certain gate configurations we find that the presence of the pump can lead to a change in sign or even to an enhancement of the PC near zero delay. For example, in Fig. 4a the PC shows saturation behavior at positive $V_{tg}$ but at $V_{tg}$ = -1 V the signal reverses polarity at zero delay, and at $V_{tg}$ = -10V the magnitude of the PC is actually enhanced by the presence of the pump. Figure 4c shows the dependence of the PC on $V_{tg}$ with the pump at zero delay (red trace) and without the pump (black trace), where the regimes of saturation, polarity reversal, and enhancement are indicated.

This behavior can be explained as resulting from the combination of photo-thermoelectric and built-in field effects. A straightforward analysis shows that at a pn junction both contributions have the same sign, and since both saturate, the pump pulse necessarily results in a reduction of the PC (see Supplementary Materials). On the other hand, at a pp or nn junction the two contributions can have opposite signs and the PC can then have either sign depending on their relative strengths. For this particular device configuration at $V_{bg}$ = -24.2 V, the photo-thermoelectric effect alone gives a positive PC. The built-in field contribution is parallel to it in the pn junction configuration ($V_t > 0$) and antiparallel in the pp junction ($V_t < 0$). Thus, the observation of the PC reversing polarity and eventually being enhanced in the pp configuration at zero pump delay implies that the photo-thermoelectric effect saturates at a lower power level than the built-in field effect. If this is so, then the factor of 10 suppression of the photocurrent on switching from CW to pulsed laser may be mainly due to the saturation of the photo-thermoelectric effect, implying that the latter is the larger contribution under CW excitation at low temperatures.

We finish by noting that at 20 K, the response time remains at about 4 ps independent of $V_{tg}$ and therefore of the junction type (Fig. 4d). This suggests that high speed graphene device can employ any kind of junction. Our results reveal several new aspects of photocurrent generation in graphene which are relevant both for fundamental understanding of non-equilibrium electron physics and for the design of ultrafast devices.

**Methods**:

**Sample Fabrication:** Graphene was mechanically exfoliated onto 285 nm thermal $SiO_2$ on highly p-doped silicon serving as a back gate. Photolithography was used to define the electrode and top-gate patterns. Titanium/gold (5nm/60 nm) was evaporated for the electrodes. The gate dielectric was 40 nm of $Al_2O_3$ grown by atomic layer deposition (ALD) at low temperature to facilitate lift-off. A large window was patterned over the whole graphene strip before ALD, after which the excess alumina was lifted off and then the gate itself was patterned. The number of layers of graphene was determined from color contrast and verified by Raman spectroscopy. The devices were wire-bonded to chip carriers and mounted in a continuous flow cryostat.

**Ultrafast Photocurrent Measurement:** Laser pulses at 800 nm were generated by a Coherent MIRA laser with 76MHz repetition rate. The laser could be switched to CW mode as needed. Autocorrelation measurements showed the pulse width is ~250fs at the sample. We used standard scanning photocurrent microscopy (SPM)[15]. The laser spot size is about 1 μm at the sample. For pump-probe measurements, the pulse is split into two optical paths, with one used as the probe for regular SPM, and the other containing a linear delay stage to vary the delay between the pump and probe pulses. A mechanical chopper running around 1.7kHz is used to modulate the probe pulse (single chopping configuration).

**Figure Captions**:

**Figure 1 | Standard photocurrent microscopy of a graphene device. a,** Experimental setup and schematic structure of a graphene device with top (gold) and bottom (dark gray) gates. **b,** (left) Optical microscope image of the graphene trilayer device, (center) scanning reflection image, and (right) photocurrent image obtained at $V_{tg}$ = 10 V, $V_{bg}$ = 0 V, and laser power of 70 µW at 20 K. Scale bar: 3 µm. **c,** Source-drain resistance as a function of $V_{tg}$ and $V_{bg}$. The dashed line indicates where charge neutrality occurs under the top gate. **d,** Photocurrent as a function of $V_{tg}$ and $V_{bg}$ with the laser fixed at the pn junction (circle in b). **e,** Photocurrent (red) and source-drain resistance (blue) as a function of $V_{tg}$ at $V_{bg}$ = 30 V.

**Figure 2 | Power dependence of pump-probe measurements. a,** Probe-induced photocurrent at a pn junction as a function of pump-probe pulse delay. The probe power is 70 µW and pump power is as indicated, with $V_{tg}$ = 7.5 V and $V_{bg}$ = -24.2 V. The solid lines are guides to the eye. **b,** Probe-induced photocurrent at zero delay (black) and response time $\tau$ (red), defined as the half-width-half-maximum of the dip in **a** as a function of pump power. **c,** Photocurrent as a function of probe power $P$ with no pump. The line is a power-law fit with $I_{pc} \sim P^{0.7}$.

**Figure 3 | Temperature dependence of photocurrent amplitude and dynamics. a,** Delay time scan of probe-induced photocurrent at 250 K (red) and 20 K (black). The probe power is 70 µW and the pump power is 270 µW, with $V_{tg}$ = 7.5 V and $V_{bg}$ = -24.2V. **b,** Temperature dependence of source drain resistance (black) and response time $\tau$ (red). **c,** Temperature dependence of photocurrent generated by CW (red) and 250-fs pulsed (black) excitation at 70 µW.

**Figure 4 | Gate dependence of pump-probe measurements implying two contributions. a,** Delay time scan of probe-induced photocurrent at a series of top gate voltages as indicated. **b,** Gate dependence of probe-induced photocurrent with (red) and without (black) pump at zero delay at $V_{bg}$ = -44.2 V. Here, the photocurrent is suppressed in the presence of the pump at all $V_{tg}$ consistent with saturation. (Probe power is 70 µW and pump power is 270 µW). **c,** As in **b** but at $V_{bg}$ = -24.2 V. Here the photocurrent can reverse polarity or be enhanced in the presence of the pump. **d,** Response time $\tau$ as a function of $V_{tg}$.

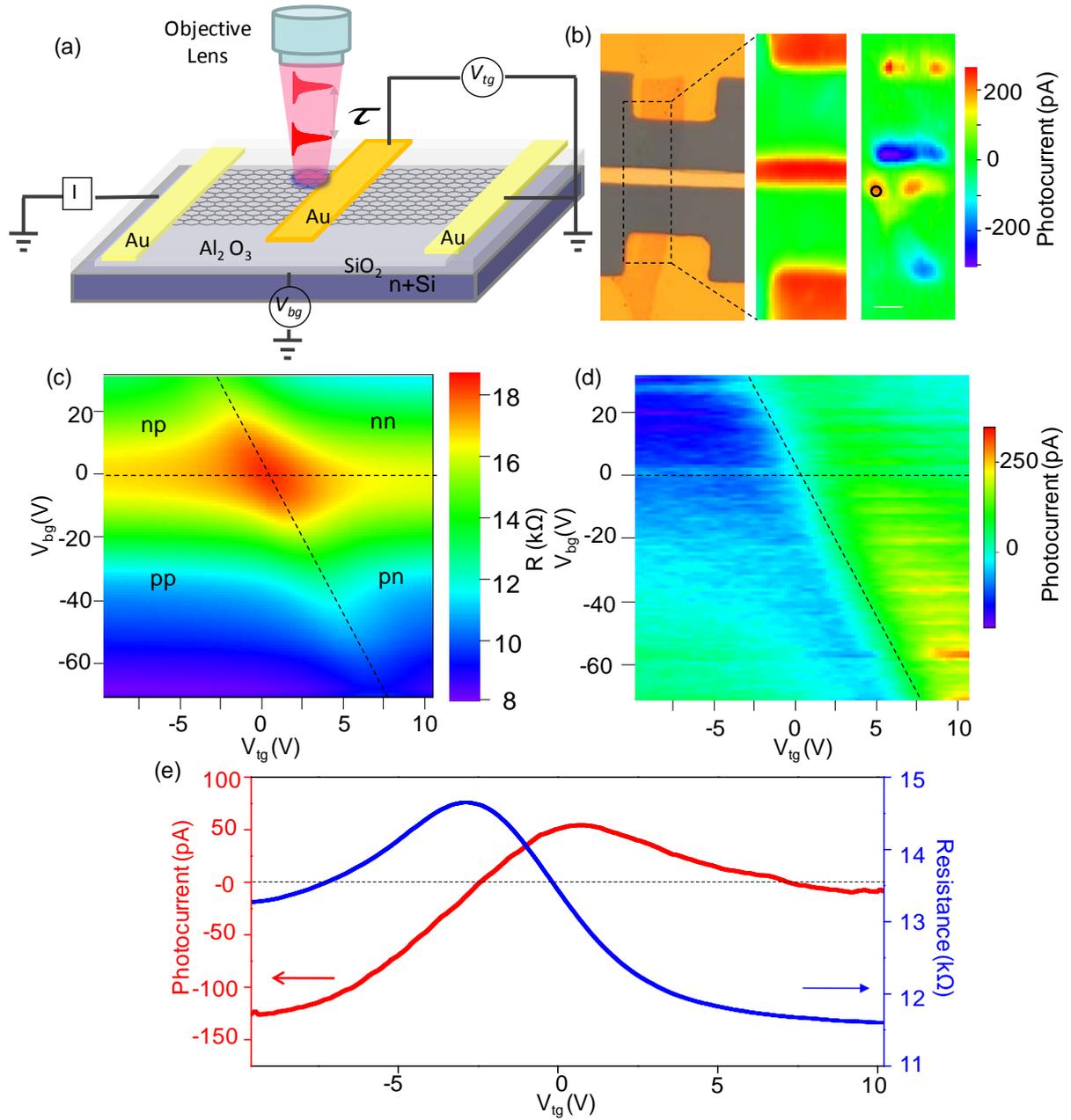

**Figure 1 | Standard photocurrent microscopy of a graphene device. a,** Experimental setup and schematic structure of a graphene device with top (gold) and bottom (dark gray) gates. **b,** (left) Optical microscope image of the graphene trilayer device, (center) scanning reflection image, and (right) photocurrent image obtained at $V_{tg}$ = 10 V, $V_{bg}$ = 0 V, and laser power of 70 µW at 20 K. Scale bar: 3 µm. **c,** Source-drain resistance as a function of $V_{tg}$ and $V_{bg}$. The dashed line indicates where charge neutrality occurs under the top gate. **d,** Photocurrent as a function of $V_{tg}$ and $V_{bg}$ with the laser fixed at the pn junction (circle in b). **e,** Photocurrent (red) and source-drain resistance (blue) as a function of $V_{tg}$ at $V_{bg}$ = 30 V.

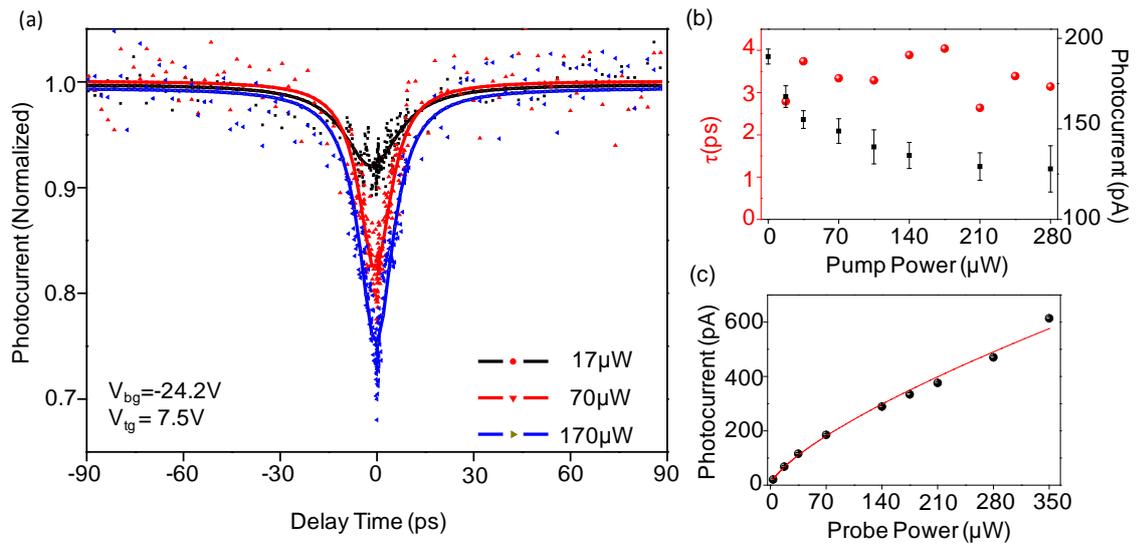

**Figure 2 | Power dependence of pump-probe measurements. a,** Probe-induced photocurrent at a pn junction as a function of pump-probe pulse delay. The probe power is 70 µW and pump power is as indicated, with $V_{tg}$ = 7.5 V and $V_{bg}$ = -24.2 V. The solid lines are guides to the eye. **b,** Probe-induced photocurrent at zero delay (black) and response time $\tau$ (red), defined as the half-width-half-maximum of the dip in **a** as a function of pump power. **c,** Photocurrent as a function of probe power $P$ with no pump. The line is a power-law fit with $I_{pc} \sim P^{0.7}$.

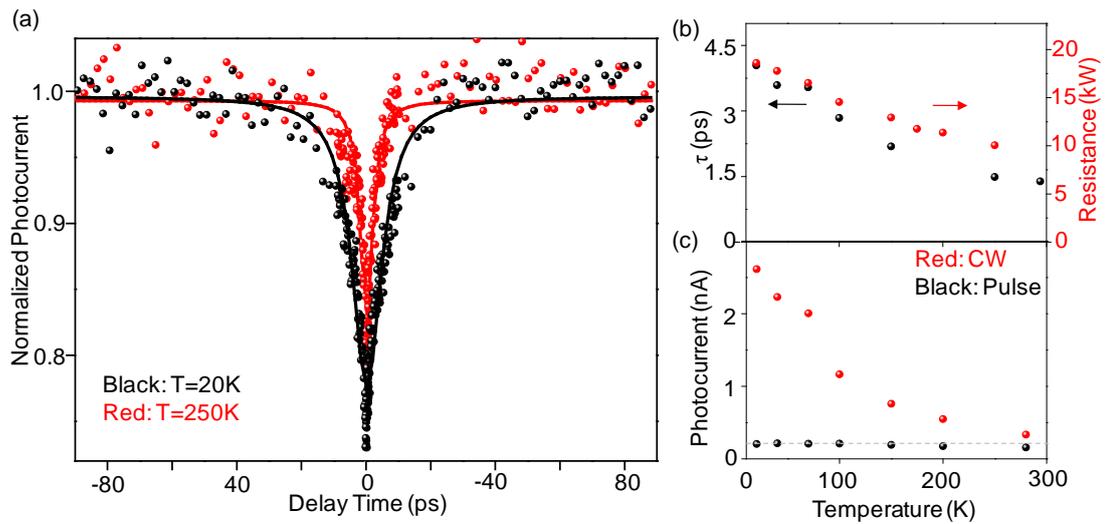

**Figure 3 | Temperature dependence of photocurrent amplitude and dynamics. a,** Delay time scan of probe-induced photocurrent at 250 K (red) and 20 K (black). The probe power is 70 μW and the pump power is 270 μW, with $V_{tg}$ = 7.5 V and $V_{bg}$ = -24.2V. **b,** Temperature dependence of source drain resistance (black) and response time τ (red). **c,** Temperature dependence of photocurrent generated by CW (red) and 250-fs pulsed (black) excitation at 70 μW.

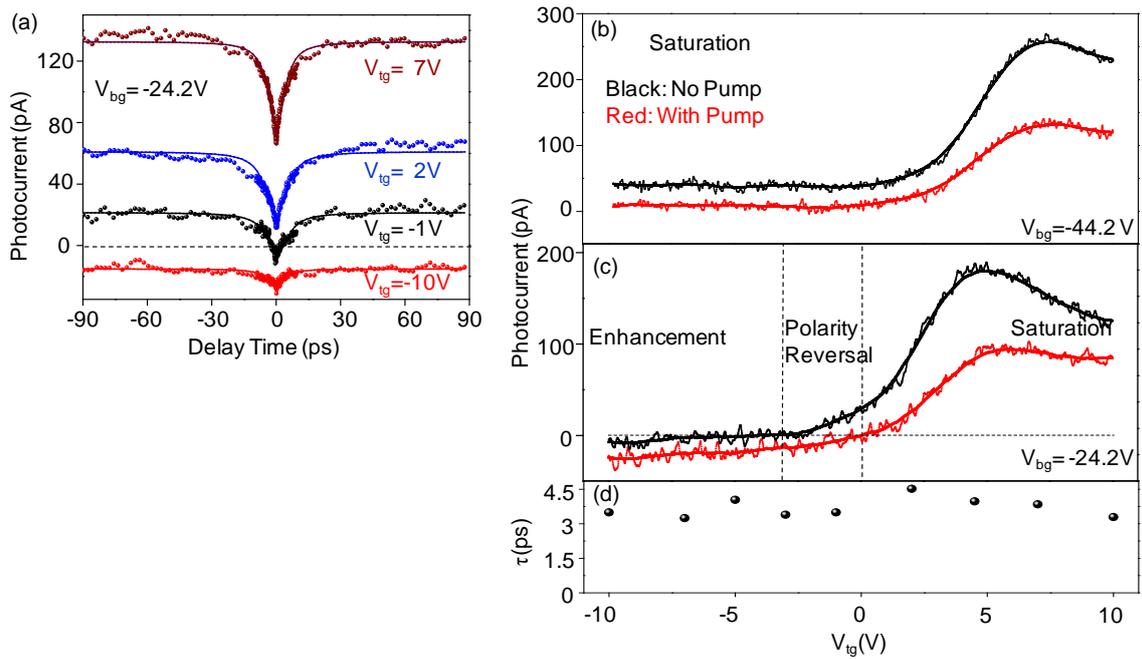

**Figure 4 | Gate dependence of pump-probe measurements implying two contributions. a,** Delay time scan of probe-induced photocurrent at a series of top gate voltages as indicated. **b,** Gate dependence of probe-induced photocurrent with (red) and without (black) pump at zero delay at $V_{bg}$ = -44.2 V. Here, the photocurrent is suppressed in the presence of the pump at all $V_{tg}$ consistent with saturation. (Probe power is 70 µW and pump power is 270 µW). **c,** As in **b** but at $V_{bg}$ = -24.2 V. Here the photocurrent can reverse polarity or be enhanced in the presence of the pump. **d,** Response time $\tau$ as a function of $V_{tg}$.

**Acknowledgments:**
We thank Zhaohui Zhong for helpful suggestions on the device fabrication and Theodore Norris for useful discussions. X. Xu acknowledges support from DARPA YFA. The research was supported in part by the US Department of Energy, Office of Basic Energy Sciences, Division of Materials Sciences and Engineering (DE-SC0002197). A. Jones was supported by the NSF Graduate Research Fellowship (DGE-0718124). Device fabrication was performed at the University of Washington Nanotechnology User Facility funded by the NSF. X. Xu thanks Blayne Heckel, Boris Blinov and Larry Sorensen for the help with the laboratory setup.

**Author Contribution:**
X.X. conceived the experiments. G.A. fabricated the devices, assisted by A.J. and D.S.. D.S. performed the measurements, assisted by X.X, G.A., and D.H.C.. W.Y. contributed to the theoretical explanation. All authors discussed the results and contributed to the paper writing.